%For Torino diquark conf 28-30 Oct, 1996
%exotic96.tex Nov 24, 1996

%\magnification=\magstep1
\font\twelvebf=cmbx10 scaled \magstep1
%\normalbaselineskip=20pt
%\baselineskip=20pt
\hsize 16.2true cm
\vsize 23.0true cm
\voffset 1.5cm
\nopagenumbers
\headline={\ifnum \pageno=1 \hfil \else\hss\tenrm\folio\hss\fi}
\pageno=1
 
\hfill IUHET--344,  IU/NTC--96--10, DFTT 59/96
 
\hfill Nov. 1996
\bigskip
\noindent{\twelvebf Diquark model of exotic mesons$^1$}
 
\medskip
\bigskip
 
\noindent D. B. Lichtenberg$^a$ and Renato Roncaglia$^b$
 
\noindent Physics Department, Indiana University, 
Bloomington, IN 47405, USA
\medskip
\noindent Enrico Predazzi$^c$ 

\noindent Dipartimento di Fisica 
Teorica, Universit\`a di Torino
 and INFN, Sezione di Torino, I-10125, Torino, Italy
 
 \vskip 1 cm

\noindent {\bf Abstract.} This is a conference mostly
devoted to diquarks. Although an exotic meson can
in principle be composed of a diquark and an antidiquark,
such an exotic is unlikely to be experimentally
observable in the near future. The reason, 
according to the model, is that diquark-antidiquark
exotics have masses well above the threshold for
decay into two mesons, and are likely to have 
widths too large to make them observable. A possible 
exception is a $ u d \bar b \bar b$ exotic, but it will
be a long time before such a state can be observed
even if it is stable against strong decay.

\bigskip
%\noindent{\bf Keywords.} Quark, diquark, meson, baryon,
%exotic, hadron
\bigskip
\vskip 6cm
$^1$Invited talk given by the first author at an international
workshop, 
Diquarks 3, Torino, Italy, October 28--30, 1996. A shortened
version will appear in the proceedings, to be published
by World Scientific, Singapore.
\medskip
$^a$lichten@indiana.edu

$^b$roncagli@indiana.edu

$^c$predazzi@to.infn.it

\vfill\eject
 
\noindent {\bf 1. Introduction}
\bigskip 
 
An exotic meson
has  a structure which is different  from  
that of a normal meson. This incomplete definition
says what an exotic is not, but does not tell what it is. 

A normal meson
has the quantum numbers of a possible bound state of a
quark and an antiquark: so-called normal quantum numbers. 
A meson which does not have normal quantum numbers 
is said to have exotic quantum numbers, and 
is by definition exotic. 
Some physicists think that
mesons with exotic quantum numbers ought to exist because
QCD does not obviously forbid them, but 
there is not yet definitive experimental
evidence for the existence of any such  meson. 

A meson may have normal quantum
numbers and still be exotic if its internal
structure differs from that of a normal meson. 
Although there are candidates for such exotics, none
has yet been positively identified.
The problem of how to distinguish between
a normal and an exotic meson with the same normal quantum
numbers is a difficult one and remains unsolved, although
progress has been made. 

Even the structure of a normal meson is not simple,
as the meson contains, in addition to a valence 
quark and antiquark, gluons and quark-antiquark pairs
of the sea. 
It is convenient to simplify the problem by considering
a model in which hadrons are made of constituent
quarks, antiquarks,  and gluons. A constituent quark
is composed of a current quark plus a cloud of gluons and
quark-antiquark pairs. The cloud has inertia, and so
the mass of a constituent quark is larger than 
the mass of a current quark.  
Empirically,  the mass difference between a constituent
quark and a current quark of the same flavour is about 300 
MeV.  A constituent gluon is composed of a massless gluon
surrounded by a cloud of gluons and pairs. Because a
gluon has a stronger colour-charge than a quark, the
inertia of its cloud is larger, and the mass of 
a constituent gluon is about 750 MeV. 

In a model with only constituent quarks and 
constituent gluons,
a normal meson is composed of a quark and antiquark. 
Among possible exotic mesons there might be glueballs
(composed of gluons only),
hybrids (composed of a quark, antiquark, and gluons), 
meson ``molecules'' (composed of two normal mesons),
and diquark-antidiquark states. The evidence for the
existence of exotic mesons is slowly accumulating, but is 
not yet definitive [1].
%Ref 1 PDG
We use the notation ``four-quark'' state or
``tetraquark'' to refer to any
exotic meson containing two quarks and two antiquarks,
whether a diquark-antiquark state or a two-meson molecule.

An unambiguous signature for an exotic meson would be the 
discovery of a meson with quantum
numbers which are impossible for a quark-antiquark pair.
Among the possibilities are mesons with
flavour quantum numbers greater than 1 or with two
different flavours (not one flavour and another antiflavour)
or mesons with
combinations of spin, parity, and charge-conjugation
parity which are forbidden to a quark-antiquark state.
Unfortunately for the prospect of easy discovery of
exotics, those with the lowest masses are likely to have
normal quantum numbers, and furthermore, are likely
to mix with ordinary quark-antiquark mesons. 
\bigskip
\noindent {\bf 2.  Rough estimate of exotic meson masses}
\bigskip

	Let us crudely estimate the constituent mass of a 
light $u$ or $d$ quark and also the constituent mass
of a gluon. We begin with the mass $m_q$ of a light $u$
or $d$ quark, neglecting their mass difference. One
way we can estimate the mass is from the proton magnetic
moment, assuming that the quarks have Dirac moments.
This procedure gives 
$$m_q = m_p/ \mu =  336\ {\rm MeV}, \eqno(2)$$
where $m_p$ is the proton mass and $\mu$ is the 
dimensionless ratio of the
proton magnetic moment to one nuclear magneton. Another
way to estimate $m_q$ is to take 1/2 the spin-averaged
mass of the $\rho$ and $\pi$ mesons. (We are 
neglecting  the meson binding
energy in the absence of spin-dependent forces). This gives
$$m_q = (3m_{\rho} + m_{\pi})/8 = 306\ {\rm MeV}. \eqno(3)$$
Still a third way is to take 1/3 the 
appropriately spin-averaged
mass of the nucleon and $\Delta$ [2], which gives 
%Ref 2 Anselm, D, Predazzi
$$m_q = (N + \Delta)/6 = 362\ {\rm MeV}. \eqno(4)$$
There is no fundamental reason for these methods to
yield the same mass, but to a rough approximation they
do. The average is 334 MeV; rounded to the nearest 10 MeV, 
it is
$$m_q = 330 \ {\rm MeV}. \eqno(5)$$

We interpret
the constituent mass of a valence quark as its current mass
plus the extra inertia of the sea of quark-antiquark
pairs and gluons surrounding it. When the quark
moves adiabatically, it drags this sea with it. 
(If we hit a quark violently, the sea
does not follow along, and we measure the current mass.)

What is the extra inertia of the sea? We can safely
neglect the current masses of the $u$ and $d$, which
are just a few MeV. Therefore, we can interpret the
quark constituent mass of about 330 MeV as 
in effect the mass of the part of the sea that is
dragged along as the quark moves adiabatically. 

We now make the conjecture that the mass of the
relevant part of the sea is proportional to the square
of the strong charge (or colour) of the quark. If so,
the relevant quantity is the Casimir operator of
SU(3), namely, $F^2$. This has the value $F_q^2 = 4/3$
for a quark. 

This little exercise enables us to estimate the mass
of a constituent gluon (if, indeed, the concept of
a constituent gluon makes sense). A gluon has 
a larger colour-charge than a quark; for a gluon, we
have $F_g^2 =3$. Therefore, 
we find that the constituent gluon mass is
$$m_g = (F_g^2 /F_q^2) m_q= (9/4) 334\ {\rm MeV}= 
750 \ {\rm MeV}, \eqno(6)$$
again rounding to the nearest 10 MeV.

We are now able to obtain  rough estimates of the
masses of exotic mesons containing light quarks and/or
glue. These estimates neglect hadron binding
energies, and, in particular, neglect the effect of 
spin-dependent forces.
Later we shall obtain estimates of some exotic masses in
specific models. 

We denote the mass of a meson by $M$,
with subscripts denoting its constituents.
A four-quark state has an estimated mass 
$$M_{qqqq}= 4 m_q = 1320\ {\rm MeV}, \eqno(7)$$ 
a hybrid has an estimated mass
$$M_{qqg}= 2m_q + m_g = 1410\ {\rm MeV},\eqno(8)$$
and a glueball made of two constituent 
gluons has an estimated mass
$$M_{gg} = 2 m_g =1500 \ {\rm MeV}. \eqno(9)$$

All these masses are in the same ballpark. Furthermore,
excited quark-antiquark (two-quark) mesons also have masses
in this region. Consequently, if quantum
numbers permit, observed mesons with masses
in the region, say, between 1300 and 1600 MeV are likely to
be mixtures of two-quark, four-quark, hybrid, and
glueball mesons. This mixing ought to occur even though
most ground-state two-quark mesons are largely unmixed.

We can also estimate the masses of exotics containing
other quark flavours. The mass differences between 
different flavoured quarks are roughly 
$$m_s- m_q =170\ -\ 200\ {\rm MeV}, \eqno(10)$$ 
$$ m_c -m_s = 1160\ -\ 1200\ {\rm MeV}, \eqno(11)$$ 
$$ m_b- m_c = 3340\ -\ 3400\ {\rm MeV}. \eqno(12)$$
Therefore, replacing a light quark in an exotic by a 
heavier one
will raise the mass of the exotic approximately by
the appropriate quark mass difference. Actually, this 
approximation is likely
to overestimate the mass of the heavier exotic because
the binding energy, which we have neglected
so far,  increases with mass. 

In order to go beyond rough estimates of exotic
masses and to calculate exotic decay rates and branching
fractions, we need to consider specific models. Some 
simple models can give precise predictions for masses 
and decay rates, but precision does not necessarily
mean correctness. Most of us believe that QCD is the correct
theory of strong interactions, but we do not know 
whether the predictions of simple models are anywhere
near the predictions of QCD.
Other models, such as lattice models of QCD are 
limited in precision by the numerical techniques 
used to evaluate them.

We concentrate on a model of exotic mesons 
composed of a diquark and an antidiquark.
We shall discuss
the strengths and weaknesses of the model, as well as
giving its predictions about exotic mesons. 
\bigskip
Work has been done with a number of other four-quark models, 
of which we mention the papers of Sylvestre-Brac and
Semay [3] and Pepin et al. [4]. Other references are
%Ref 3 Ref 4
contained in these papers and in a review of diquarks [5].
%Ref 5 
\bigskip
\noindent {\bf 3. A diquark model}
\bigskip

In our scheme, an exotic hadron is composed of 
quark clusters [6], and we confine ourselves to the
%Ref 6 DBL Wein	
case in which the clusters are diquarks or antidiquarks.
Now  the ${\bf F}_1 \cdot
{\bf F}_2$ factor in the lowest-order QCD potential between
two quarks is positive for a sextet diquark but 
negative for a triplet (actually antitriplet) diquark, 
leading to a
repulsive interaction between two quarks at small
separations for the former and an attractive 
interaction for the latter. 
Therefore, we expect that colour-sextet diquarks will
lie higher in energy than the colour-triplet
diquarks. While we neglect the
former, on general grounds we expect that their 
effect could slightly lower the ground state energy
compared to our finding. We do not, however, have a
simple recipe to propose in order to give a quantitative
estimate of this effect.

We now consider a detailed model which allows
us to obtain the masses of exotic mesons in terms of
the masses of mesons and baryons. Many of the
 meson and baryon masses are known either from experiment
or from predictions for unknown hadrons 
based on the systematics of known hadrons [7,8].
We can use this information as input data to
obtain specific predictions for the
masses of exotics. Our model has the advantage that
we do not need to assume an explicit Hamiltonian
to describe the interaction, but there is the
accompanying disadvantage that the model does not
allow us to calculate decay rates. We can, however,
make some qualitative remarks about decays.

According to QCD, in first approximation, the force
between two coloured particles depends only on their
colour configuration and not on their mass or spin.
This fact gives rise to an approximate supersymmetry
between a diquark and antiquark. A discussion of this
supersymmetry, with references, is contained in 
the diquark review  [5]. 
%Ref 5 

We neglect spin at the outset of our treatment, but
subsequently take it into account. 
However, we cannot neglect the effect of mass because,
even if the force does not depend on the constituent
masses, the meson mass does. 
The mass of a composite particle is the sum of the
masses of the constituents plus an interaction energy.
In general, this interaction energy also depends on the 
constituent masses even if the force does not.

In our approximation, the force between a quark 
and an antiquark in an overall
colour-singlet state is the same as the force 
between an antidiquark and a 
diquark in an overall colour singlet. Consequently, we can
equate the mass of an exotic meson composed of a diquark
and antidiquark to the mass of a meson composed of a
fictitious quark and antiquark having the same masses as the
diquark and antidiquark. 

Two obstacles must be overcome in order to carry out
this procedure:  we must obtain estimates of the
masses of diquarks, and we must obtain the
interaction energy of two fictitious quarks having the same 
masses as the diquarks.

Let us first consider the problem of obtaining 
the interaction energy of a bound quark and antiquark
with any masses. The interaction
energy turns out to be a fairly
smooth function of the reduced 
mass of the two particles [7]. We 
%Ref 7 Renato
assign reasonable values of the masses to the different
flavoured quarks. To be specific, we use the same
quark masses as in ref.\ [8], which are (in MeV):
%Ref 8 Renato
$$m_q = 300, \quad m_s = 475, \quad m_c = 1640,
\quad m_b = 4985. \eqno(13). $$
These masses are a little below those given in our
rough estimates in Eqs.\ (5) and (10)--(12). 
This is not important,
as we have found [7,8] that we may vary 
the input quark masses appreciably without significantly
changing the ouput hadron masses, as we can make
compensating changes in the interaction energy. 
Our approach [7,8] is much more sensitive to the differences
of quark input masses than to their absolute values.

If we use the experimental values $M_{12}$ of meson 
masses [1] as input,
we can calculate the interaction energies $E_{12}$ from the
simple formula:
$$E_{12}= M_{12} - m_1 - m_2. \eqno(14) $$
In order to eliminate the effects of spin as much
as possible in this procedure, we use spin-averaged
meson and baryon masses according to the prescription
of Ref.\ [2]. This procedure cannot always be carried
out in terms of known hadron masses. Where necessary,
we approximately remove the effect of the colour-magnetic
interaction by a semi-empirical mass formula [9].
%Ref 9 Wang

We can then plot the values of $E_{12}$ against the
reduced mass $\mu_{12}$. We obtain the value of $E_{12}$
for any $\mu_{12}$ by interpolation or extrapolation,
or, in other words, from the mass given by a smooth
curve connecting the known points. 
Thus, if we are given the masses of any two diquarks, we
compute the reduced mass, find 
the relevant interaction energy from the curve
for mesons, add the two diquark masses, and thereby
obtain the mass of the exotic composed of the two
diquarks. 

We next turn to the problem of estimating the diquark masses.
Consider the spin average of the masses of ground-state
baryons with a given quark content. The individual baryon
masses are known either from  
experiment [1] or from estimates based on the systematics
of known baryons  [7,8]. For a baryon like the $\Omega$,
which contains the quarks $sss$, spin averaging in terms
of known baryons cannot be done. In such cases we use
the semi-empirical formula [9]. 

We assume the baryon is composed of a diquark
and a quark. If the quarks have different flavours, then
we assume that the two heaviest quarks form the diquark.
(We neglect the mass difference between the $d$ and $u$ 
quarks.)
We then guess a trial value for the diquark mass; it
is convenient to take the sum of the masses of the two
quarks in the diquark.
We then find the reduced mass of the diquark
and the third (or spectator) quark, use the meson curve
to obtain the interaction energy, and so obtain a 
prediction for the baryon mass, which,
in general, will not be the same as the input
baryon mass. However, we can then use a revised 
diquark mass and repeat the process by iteration,
stopping when we have found an input diquark mass
that gives the input value of the baryon mass.
The difference in mass between a diquark and the
sum of the masses of the quarks it contains is the
triplet interaction energy $E^t_{12}$. We expect that
the energy
$E^t_{12}$ is larger than $E_{12}$ because the interaction
at small distances contains the factor ${\bf F}_1 \cdot
{\bf F}_2$, which is smaller in magnitude for two
quarks in a colour-triplet (actually, antitriplet) state
than in a colour-singlet state. We find numerically
that our expectation is satisfied.

We must overcome still another difficulty. If we use
the above iteration procedure to calculate the mass of,
say, the $ss$ diquark, the result depends mildly on
whether the spectator quark in the baryon is  $q$
($u$ or $d$) or $s$. We resolve the ambiguity
by  averaging the two values.

The above procedure gives us  diquark masses that
are suitable for use 
in calculating the masses of exotic mesons except that
so far we have not included the effects of spin.
We therefore have to correct the diquark masses for 
spin-dependent forces. We do so under the assumption
that these forces arise from the colour-magnetic
interaction of perturbative QCD or from a generalization
of this interaction [2, 7, 8]. Then,
any pair of quarks in a baryon is subject to 
essentially the same spin-dependent force as the
same two quarks in an exotic meson, provided the
two quarks belong to a single diquark. 

The magnitude
of the colour-magnetic energy arising from two quarks
in a baryon has already
been estimated [2, 7, 8] and found to be only weakly
dependent on the mass of the 
spectator quark. This dependence
is such that the magnitude of the colour-magnetic
energy increases slightly as the mass of the third 
quark increases. We approximate the colour-magnetic
energy by taking an average of the energy obtained
with the $q$, $s$, and $c$ as spectator. (We do not
include the $b$ as spectator 
because the result is very similar to the $c$ with
a larger error and because we do not want to give
too much importance to the heavy quarks.) A different
prescription will change our results by only a few MeV
at most.  We can now obtain all the necessary colour-magnetic
energies from the results in Refs.\ [2,7,8]. 

The above procedure lets us correct the masses of the
diquarks for the effects of spin. However, in general,
in an exotic meson
there are still other spin-dependent forces arising
between the quarks in the diquark and the quarks in 
the antidiquark. 
However, if either
the diquark or antidiquark has spin zero, the net
effects of these spin-dependent inter-diquark  forces 
vanish. Although we have a method that applies to
exotics containing two spin-one diquarks, here we 
restrict ourselves for simplicity to exotics 
with at least one diquark of spin zero. These latter
exotics have the lowest masses in any case.

\bigskip
\noindent {\bf 4. Results and discussion}
\bigskip

As we stated in the previous section, we obtain diquark
masses in terms of input baryon masses and estimates
of the colour-magnetic interaction energies between
pairs of quarks in baryons. Strictly speaking, the
calculated diquark masses depend slightly on the
mass of the spectator quark in the baryon. The differences,
however, are usually less than 20 MeV. In those cases in 
which we
obtain more than one value of the diquark mass,  we give
the average value. These average  values of the diquark
masses are shown in Table 1. 

\bigskip
%\vfill\eject

Table 1. Masses  $M_0$ and $M_1$ of spin-zero and spin-one
diquarks obtained from input baryon masses and 
colour-magnetic interaction energies in baryons. The method
is described in more detail in Sec.\ 3. The symbol $q$
stands for $u$ or $d$, and masses are
rounded to the nearest 5 MeV. Errors of up to 20 MeV
arise because, in our model, the calculated diquark masses 
depend on the spectator quark, and we have averaged over this
dependence.

\vskip 12pt
$$\vbox {\halign {\hfil #\hfil &&\quad \hfil #\hfil \cr
\cr \noalign{\hrule}%\cr
\cr \noalign{\hrule}
\cr
Quark content & $M_0$ (MeV)& $M_1$ (MeV) \cr 
\cr \noalign{\hrule}
\cr
$qq$ &   595 &   800   \cr
$qs$ &   835 &   975   \cr
$ss$ &     - &  1150   \cr
$qc$ &  2100 &  2150   \cr
$sc$ &  2250 &  2295   \cr
$cc$ &  -    &  3415   \cr
$qb$ &  5465 &  5485   \cr
$sb$ &  5630 &  5650   \cr
$cb$ &  6735 &  6750   \cr
$bb$ & -     & 10075   \cr
 
\cr \noalign{\hrule}%\cr
\cr \noalign{\hrule}
}}$$
 
%$^a$ 
\bigskip

Using the diquark masses in Table 1, we can compute
the reduced mass of a system of a diquark and antidiquark.
We then obtain the interaction energy for the reduced
mass, using the meson curve as input. This procedure
gives us the exotic meson mass. We have already
taken into account the colour-magnetic
interaction in computing the diquark masses. Because
we restrict ourselves to those cases in which at 
least one diquark has spin zero, no further spin-dependent
forces enter the problem. The calculated exotic masses
are shown in Table 2. We also give in Table 2 the
lightest mesons into which the exotic can decay strongly,
provided the energy permits. The sum of the masses of
the decay products are shown in the last column of Table 2.

The calculated exotic masses given in Table 2 have errors
of up to 30 MeV associated with the fact that we have
chosen the fixed diquark masses of Table 1, whereas 
actually, in our model, the diquark masses depend on their
environment. If future experiments should yield masses
which differ from those in Table 2 by much more than 30
MeV, we will have to discard our model in its present form.

We see from Table 2 that only the exotic $qq\bar b\bar b$
lies below the threshold of the lightest mesons into
which it can decay strongly. The exotic 
$qs\bar b\bar b$ lies only a little above threshold.
It may be observable as a narrow resonance or even a
bound state, as our model is probably not accurate
enough to distinguish these possibilities in this
case. However,
exotic mesons with two $b$ quarks are unlikely to
be seen for quite some time.

%Don. Added paragraph 
The exotic masses we have calculated depend on the input 
masses of mesons and baryons. Because some of these input 
masses are not known from experiment but are estimated,
we have an additional source of error. In particular, 
the estimation of the masses of baryons containing two 
heavy quarks involves a considerable extrapolation 
from known data [8]. This fact adds to the uncertainty in
our calculated masses of exotics containing two heavy quarks.

All exotics with at most one heavy quark have masses 
well above
the masses of the decay products, as we see from
Table 2.  This means that these
exotics will decay rapidly into two mesons. 
Our method does not allow us to calculate the decay
widths (because we do not have a Hamiltonian or 
spatial wave functions). However, we observe that the 
colour wave functions of the exotics are not orthogonal
to the colour wave functions of the two mesons into
which they can decay. This means that there is nothing
to prevent the exotics from ``falling apart'' into their
decay products. Under these circumstances, we guess
that the exotics will have widths too large to make
these mesons observable.

It is possible that the spin-dependent forces arise,
not from the colour-magnetic interaction but from
pseudoscalar meson exchange [10, 11]. If so, 
%Ref 10 11 Glozman
some exotic mesons with two $c$ quarks
are predicted [4] to have  masses  well below those
given in our Table 2 and to be stable against strong
decay. Future experiments, by their observation or
non-observation of weakly
decaying exotics containing two $c$ quarks,
will rule out either our model or the one of Ref.\ [4].

We should like to thank Ted Barnes, Michael Pennington,
and Jean-Marc Richard
for valuable discussions. This work was supported in
part by the U.S. Department of Energy, the U.S.
National Science Foundation, by the Italian
Institute for Nuclear Physics (INFN) and
by the Ministry of Universities, Research,
Science and Technology (MURST) of Italy.

\bigskip
\noindent References
\medskip

[1] Particle Data Group: R. M. Barnett et al.,
Phys.\ Rev.\ D 54 (1996) 1.

[2] M.  Anselmino,
D.B. Lichtenberg, and E. Predazzi, Z.\ Phys.\ C. {\bf 48},
(1990) 605.
 
[3] B. Silvestre-Brac and C. Semay, Z.\ Phys.\ C {\bf 59}
(1993) 457.
 
[4] S. Pepin, Fl.\ Stancu, M. Genovese, and J.-M.\
Richard, nucl-th/9608058 and ph-9609348 (unpublished).

[5] M. Anselmino, E. Predazzi, S. Ekelin,              
S. Fredriksson, and D.B. Lichtenberg, Rev.\ Mod.\ Phys.\
{\bf 65} (1993) 1199.
 
[6] D.B. Lichtenberg, E. Predazzi, D.H. Weingarten,
and J.G. Wills, Phys.\ Rev.\ D {\bf 18} (1978) 2569.

[7] R. Roncaglia,  A.R. Dzierba, D.B. Lichtenberg,   
and E. Predazzi, Phys.\ Rev.\ D {\bf 51} (1995) 1248.

[8] R. Roncaglia, D.B. Lichtenberg,                   
and E. Predazzi, Phys.\ Rev.\ D {\bf 52} (1995) 1722.
 
[9] Yong Wang and D.B. Lichtenberg, Phys.\ Rev.\ D {\bf 42}
(1990) 2404.
 
[10] L. Ya.\ Glozman and D.O. Riska, Phys.\ Rep.\
{\bf 208}, 263 (1996).
 
[11] L. Ya.\ Glozman, Z. Papp, and W. Plessas,
Phys.\ Lett. B {\bf 381} (1996) 311.

\bigskip
\vfill\eject

Table 2. Predicted masses $M_E$ of exotic
mesons  obtained from the diquark masses in Table
1 by the method described in Sec.\ 3.
These are all ground-state mesons with the given spin and quark
content, and the parities are all positive.
The symbol $q$ stands for $u$ or $d$, and
masses are rounded to the nearest 10 MeV.
The two numbers in column 2 are the spin of the diquark
and antidiquark respectively; the spin of the exotic
is just the sum of these spins. 
The next-to-last column gives the lowest-mass
mesons into which the
exotic can decay strongly if energetically permitted, 
and the last column gives the threshold energy $E_t$ 
of the decay. Errors of up to 30 MeV in the exotic
masses arise because of errors in the diquark masses.

%Renato: whenever an $s \bar s$ pair could be formed in the
%        decay products, I have calculated the threshhold
%        by using the \eta (mass assumed as 547 MeV). I have
%        also calculated the lowest decay that doesn't 
%        include any \eta, since this is not pure $s \bar s$
%        state.
 
\vskip 12pt
$$\vbox {\halign {\hfil #\hfil &&\quad \hfil #\hfil \cr
\cr \noalign{\hrule}%\cr
\cr \noalign{\hrule}
\cr
Quark content & Diquark spins & $M_E$ (MeV) &  Decay products
& $E_t$ (MeV)\cr
\cr \noalign{\hrule}
\cr
$qq\bar q\bar q$ & 0, 0 & 1180 & $\pi\pi$    &  280  \cr
$qq\bar q\bar q$ & 0, 1 & 1370 & $\pi\rho$   &  910  \cr

$qq\bar q\bar s$ & 0, 0 & 1400 & $\pi K$     &  630  \cr
$qq\bar q\bar s$ & 0, 1 & 1530 & $\pi K^*$   & 1030 \cr %Renato: was 630
$qq\bar q\bar s$ & 1, 0 & 1580 & $\pi K^*$   & 1030 \cr %Renato: was 630

$qq\bar s\bar s$ & 0, 1 & 1700 & $K K^*$     & 1390 \cr

$qs\bar q\bar s$ & 0, 0 & 1610 & $K \bar K$  &  990 \cr
$qs\bar q\bar s$ & 0, 1 & 1740 & $K^* \bar K$, $K\bar K^*$ & 1390 \cr

$qq\bar q\bar c$ & 0, 0 & 2620 & $\pi \bar D$   &  2010 \cr
$qq\bar q\bar c$ & 0, 1 & 2660 & $\pi \bar D^*$ &  2150 \cr
$qq\bar q\bar c$ & 1, 0 & 2770 & $\pi \bar D^*$ &  2150 \cr

$qq\bar q\bar b$ & 0, 0 & 5950 & $\pi B$   &  5420 \cr
$qq\bar q\bar b$ & 0, 1 & 5970 & $\pi B^*$ &  5460 \cr
$qq\bar q\bar b$ & 1, 0 & 6100 & $\pi B^*$ &  5460 \cr

$qs\bar s\bar s$ & 0, 1 & 1900 & $K \phi $     & 1520 \cr
                 &      &      & $K^* \eta$    & 1440 \cr

$qq\bar s\bar c$ & 0, 0 & 2760 & $K \bar D$   &  2360 \cr
$qq\bar s\bar c$ & 0, 1 & 2810 & $K \bar D^*$ &  2500 \cr
$qq\bar s\bar c$ & 1, 0 & 2920 & $K \bar D^*$ &  2500 \cr

$qs\bar q\bar c$ & 0, 0 & 2800 & $\pi \bar D_s$   & 2110 \cr
$qs\bar q\bar c$ & 0, 1 & 2850 & $\pi \bar D^*_s$ & 2250 \cr
$qs\bar q\bar c$ & 1, 0 & 2920 & $\pi \bar D^*_s$ & 2250 \cr

$qs\bar s\bar c$ & 0, 0 & 2950 & $K \bar D_s$   & 2360 \cr
                 &      &      & $\eta \bar D$  & 2020 \cr
$qs\bar s\bar c$ & 0, 1 & 2990 & $K \bar D^*_s$ & 2610 \cr
$qs\bar s\bar c$ & 1, 0 & 3060 & $K \bar D^*_s$ & 2610 \cr

$qc\bar s\bar s$ & 0, 1 & 3060 & $K D^*_s$     & 2610 \cr

$ss\bar s\bar c$ & 1, 0 & 3210 & $\phi D_s$   & 2990 \cr
                 &      &      & $\eta D^*_s$ & 2660 \cr

$qq\bar s\bar b$ & 0, 0 & 6120 & $K \bar B$   &  5780 \cr
$qq\bar s\bar b$ & 0, 1 & 6140 & $K \bar B^*$ &  5910 \cr
$qq\bar s\bar b$ & 1, 0 & 6260 & $K \bar B^*$ &  5910 \cr

$qs\bar q\bar b$ & 0, 0 & 6120 & $\pi \bar B_s$   & 5510 \cr
$qs\bar q\bar b$ & 0, 1 & 6140 & $\pi \bar B^*_s$ & 5550 \cr
$qs\bar q\bar b$ & 1, 0 & 6230 & $\pi \bar B^*_s$ & 5550 \cr

$qs\bar s\bar b$ & 0, 0 & 6280 & $K \bar B_s$   & 5860 \cr
                 &      &      & $\eta \bar B$  & 5830 \cr
$qs\bar s\bar b$ & 0, 1 & 6300 & $K \bar B^*_s$ & 5910 \cr
$qs\bar s\bar b$ & 1, 0 & 6390 & $K \bar B^*_s$ & 5910 \cr
                 &      &      & $\eta \bar B^*$& 5870 \cr
\cr \noalign{\hrule}%\cr
\cr \noalign{\hrule}
}}$$
 
\vfill\eject

Table 2 continued.

\vskip 12pt
$$\vbox {\halign {\hfil #\hfil &&\quad \hfil #\hfil \cr
\cr \noalign{\hrule}%\cr
\cr \noalign{\hrule}
\cr
Quark content & Diquark spins & $M_E$ (MeV) &  Decay products
& $E_t$ (MeV)\cr
\cr \noalign{\hrule}
\cr
$qb\bar s\bar s$ & 0, 1 & 6360 & $K B^*_s$     & 5910 \cr

$ss\bar s\bar b$ & 1, 0 & 6530 & $\phi B_s$   & 6390 \cr
                 &      &      & $\eta B^*_s$ & 5960 \cr

$qq\bar c\bar c$ & 0, 1 & 3910 & $\bar D \bar D^*$ & 3880 \cr

$qc\bar q\bar c$ & 0, 0 & 3920 & $\pi \eta_c$ & 3120 \cr
$qc\bar q\bar c$ & 0, 1 & 3970 & $\pi J/\psi$ & 3240 \cr

$qs\bar c\bar c$ & 0, 1 & 4090 & $\bar D \bar D^*_s$ & 4010 \cr

$qc\bar s\bar c$ & 0, 0 & 4060 & $K \eta_c$ & 3470 \cr
$qc\bar s\bar c$ & 0, 1 & 4100 & $K J/\psi$ & 3590 \cr
$qc\bar s\bar c$ & 1, 0 & 4110 & $K J/\psi$ & 3590 \cr

$sc\bar s\bar c$ & 0, 0 & 4200 & $D_s \bar D_s$ & 3940 \cr
                 &      &      & $\eta \eta_c$ & 3530 \cr
$sc\bar s\bar c$ & 0, 1 & 4240 & $D_s \bar D^*_s, D^*_s \bar D_s$ & 4080 \cr
                 &      &      & $\eta J/\psi$ & 3650 \cr

$qq\bar c\bar b$ & 0, 0 & 7220 & $\bar D \bar B$   & 7150 \cr
$qq\bar c\bar b$ & 0, 1 & 7230 & $\bar D \bar B^*$ & 7190 \cr
$qq\bar c\bar b$ & 1, 0 & 7360 & $\bar D \bar B^*$ & 7190 \cr

$qc\bar q\bar b$ & 0, 0 & 7180 & $\pi \bar B_c$   & 6390 \cr
$qc\bar q\bar b$ & 0, 1 & 7200 & $\pi \bar B^*_c$ & 6460 \cr
$qc\bar q\bar b$ & 1, 0 & 7220 & $\pi \bar B^*_c$ & 6460 \cr

$qs\bar c\bar b$ & 0, 0 & 7380 & $\bar D \bar B_s$ & 7240 \cr
$qs\bar c\bar b$ & 0, 1 & 7400 & $\bar D \bar B^*_s$ & 7280 \cr
$qs\bar c\bar b$ & 1, 0 & 7490 & $\bar D \bar B^*_s$ & 7280 \cr

$qc\bar s\bar b$ & 0, 0 & 7340 & $K \bar B_c$   & 6750 \cr
$qc\bar s\bar b$ & 0, 1 & 7360 & $K \bar B^*_c$ & 6820 \cr
$qc\bar s\bar b$ & 1, 0 & 7380 & $K \bar B^*_c$ & 6820 \cr

$qb\bar s\bar c$ & 0, 0 & 7310 & $K B_c$ & 6750 \cr
$qb\bar s\bar c$ & 0, 1 & 7350 & $K B^*_c$ & 6820 \cr
$qb\bar s\bar c$ & 1, 0 & 7330 & $K B^*_c$ & 6820 \cr

$ss\bar c\bar b$ & 1, 0 & 7620 & $\bar D_s \bar B^*_s$ & 7380 \cr

$sc\bar s\bar b$ & 0, 0 & 7470 & $\bar B_s D_s$ & 7300 \cr
                 &      &      & $\eta \bar B_c$ & 6810 \cr
$sc\bar s\bar b$ & 0, 1 & 7490 & $\bar B^*_s D_s$ & 7340 \cr
                 &      &      & $\eta \bar B^*_c$ & 6870 \cr
$sc\bar s\bar b$ & 1, 0 & 7520 & $\bar B^*_s D_s$ & 7340 \cr
                 &      &      & $\eta \bar B^*_c$ & 6870 \cr

$qq\bar b\bar b$ & 0, 1 & 10550 & $\bar B \bar B^*$ & 10600 \cr

$qb\bar q\bar b$ & 0, 0 & 10380 & $\pi \eta_b$   & 9540 \cr
$qb\bar q\bar b$ & 0, 1 & 10400 & $\pi \Upsilon$ & 9600 \cr

$qs\bar b\bar b$ & 0, 1 & 10710 & $\bar B \bar B^*_s$ & 10700 \cr

$qb\bar s\bar b$ & 0, 0 & 10550 & $K \eta_b$   & 9900 \cr
$qb\bar s\bar b$ & 0, 1 & 10570 & $K \Upsilon$ & 9960 \cr
$qb\bar s\bar b$ & 1, 0 & 10570 & $K \Upsilon$ & 9960 \cr

$sb\bar s\bar b$ & 0, 0 & 10710 & $\bar B_s B_s$ & 10740 \cr
                 &      &       & $\eta \eta_b$ & 9900 \cr
$sb\bar s\bar b$ & 0, 1 & 10730 & $\bar B^*_s B_s$ & 10780 \cr
                 &      &       & $\eta \Upsilon$ & 9950 \cr

\cr \noalign{\hrule}%\cr
\cr \noalign{\hrule}
}}$$

\bye